\documentclass[prl,twocolumn,amsmath,amssymb,floatfix]{revtex4}
\usepackage{graphicx}
\usepackage{amsmath}
\usepackage{color}
\usepackage{epstopdf}
\usepackage{multirow}
\usepackage[colorlinks=true,citecolor=blue,linkcolor=magenta]{hyperref}

\begin{document}

\title{Dephasing-insensitive quantum information storage and processing with superconducting qubits}

\author{Qiujiang Guo$^{1}$}
\author{Shi-Biao Zheng$^2$}
\email{t96034@fzu.edu.cn}
\author{Jianwen Wang$^3$, Chao Song$^1$, Pengfei Zhang$^1$, Kemin Li$^1$, Wuxin Liu$^1$, \mbox{Hui Deng}$^4$, Keqiang Huang$^{4,5}$, Dongning Zheng$^{4,5}$}
\author{Xiaobo Zhu$^{3,6}$}
\email{xbzhu16@ustc.edu.cn}
\author{H. Wang$^{1,3}$}
\email{hhwang@zju.edu.cn}
\author{C.-Y. Lu$^{3,6}$}
\author{Jian-Wei Pan$^{3,6}$}
\affiliation{$^1$ \mbox{Department of Physics, Zhejiang University, Hangzhou, Zhejiang 310027, China},
$^2$ Fujian Key Laboratory of Quantum Information and Quantum Optics, \mbox{College of Physics and Information Engineering, Fuzhou University, Fuzhou, Fujian 350116, China},
$^3$ CAS Center for Excellence and Synergetic Innovation Centre in Quantum Information and Quantum Physics, University of Science and Technology of China, Hefei, Anhui 230026, China,
$^4$ \mbox{Institute of Physics, Chinese Academy of Sciences, Beijing 100190, China},
$^5$ \mbox{School of Physical Sciences, University of Chinese Academy of Sciences, Beijing 100049, China},
$^6$ Shanghai Branch, National Laboratory for Physical Sciences at Microscale and Department of Modern Physics, University of Science and Technology of China, Shanghai 201315, China
}

\date{\today }
\bibliographystyle{apsrev4-1}


\begin{abstract}
A central task towards building a practical quantum computer is to protect individual qubits from decoherence 
while retaining the ability to perform high-fidelity entangling gates involving 
arbitrary two qubits. Here we propose and demonstrate a dephasing-insensitive 
procedure for storing and processing quantum information in an all-to-all connected superconducting circuit
involving multiple frequency-tunable qubits, each of which can be controllably coupled to any other through 
a central bus resonator. Although it is generally believed that the extra frequency tunability
enhances the control freedom but induces more dephasing impact for superconducting qubits, 
our results show that any individual qubit can be dynamically decoupled 
from dephasing noise by applying a weak continuous and resonant driving field whose phase 
is reversed in the middle of the pulse. More importantly, 
we demonstrate a new method for realizing two-qubit phase gate with inherent dynamical decoupling 
via the combination of continuous driving and qubit-qubit swapping coupling.
We find that the weak continuous driving fields not only enable the conditional dynamics 
essential for quantum information processing, 
but also protect both qubits from dephasing during the gate operation.
\end{abstract}

\vskip 0.5cm
\maketitle

\narrowtext

The advantage of quantum computation is based on
the ability of storing and processing information encoded in a collection of
qubits, which are in superposition states. An obstacle against
implementation of a complex quantum algorithm is the decoherence effect,
arising from the inevitable interaction between the quantum machine and its
environment. 
Protecting quantum information from decoherence is therefore
essential for realization of a practical quantum computational task; the
strategy depends on the properties of the noise. When the qubits are
identically coupled to the environment, decoherence due to random phase
errors can be suppressed by storing the quantum information in a
decoherence-free subspace--encoding a logic qubit into two physical qubits~\cite{Zarnadi1997PRL, Lidar1998PRL, Duan1998PRA}. 
With a suitable choice of the logic basis states, any superposition of
these states remains invariant under collective dephasing, corresponding to
a decoherence-free state.

When the noises on different qubits are uncorrelated, there does not exist such 
decoherence-free subspace; this is the case for superconducting circuits,
where the noises are localized around individual physical qubits. 
In this case, an effective strategy to suppress dephasing is
dynamical decoupling (DD), achieved by periodically applying a series of pulses
to the quantum system to refocus the system-environment evolution~\cite{Viola1998PRA, Khodjasteh2005PRL, Uhrig2007PRL}.
Recent experiments have demonstrated significant improvement of the coherence times of quantum
memories resulting from the pulsed DD~\cite{JFDu2009nature, Biercuk2009nature, Damodarakurup2009PRL, Lange2010science, Ryan2010PRL, Souza2011PRL, Naydenov2011PRB, Bylander2011nphys, Sar2012nature}. 
However, implementation of this method may be challenging for systems with fast
fluctuating noise, since it requires that the delay between the pulses be
shorter than the correlation time of the fluctuating environment. Another problem is how
to combine the decoupling scheme with gate operations, so that dephasing
is suppressed during both the information storage and processing.

An alternative way for DD is to replace pulse sequences
with continuous driving fields, which is not subject to the physical
restriction associated with the pulsed approach \cite{Torrey1949PR, Fanchini2007PRA, JMCai2012NJP}. 
Another benefit of using continuous driving is that DD can be easily
incorporated into two-qubit logic operations, as theoretically proposed \cite{Zheng2002PRA, Bermudez2012PRA}
and experimentally demonstrated with trapped ions \cite{TRTan2013PRL}. 
However, the integration of DD sequences with quantum gates in a scalable architecture like superconducting circuits
remains a nontrivial open problem~\cite{Paladino2014}. Two types of entangling gates with superconducting qubits,
the adiabatic controlled phase gate~\cite{Barends2014nature} and the cross-resonance gate~\cite{Sheldon2016}, 
have been demonstrated with fidelity values approaching the fault-tolerant threshold. 
But both gates do not take advantage of DD, and are only appropriate to neighboring qubits
with direct couplings in a circuit with limited connectivity, 
which could significantly raise the algorithmic complexity~\cite{McKay2017}.
Here we propose and demonstrate a dephasing-insensitive quantum computation scheme in an
all-to-all connected superconducting circuit featuring multiple frequency-tunable qubits connected
by the central bus resonator, where a weak microwave drive dresses each qubit
and protects it from dephasing during both the information storage and processing. 
The advantage of the frequency tunability is that the couplings between qubits 
can be dynamically switched on and off by tuning these qubits on- and off-resonance, respectively.
However, such tunability enabled by the on-chip flux coil also incurs more flux noise 
when the qubit is tuned away from its sweetpoint, i.e., the maximum frequency.
It is in this scenario that our scheme comes into play with a huge benefit.
With the application of the driving field, the observed pure dephasing
time is prolonged significantly compared with the spin echo method.
For implementation of the two-qubit entangling gate, the 
microwave drives not only help to realize the conditional dynamics without
employing non-computational states, but also protect the operation from dephasing. 
The scheme works no matter whether the qubits are coupled
through a resonator~\cite{Paik2016} or capacitor/inductor, and high-fidelity gates are promising
with further improvements in the device design.

We first show how a qubit, with its bare upper and lower energy levels denoted 
as $\left| 1\right\rangle $ and $\left| 0\right\rangle $, respectively, can be protected from dephasing under a continuous
and resonant driving field. In the frame rotating at the qubit frequency,
the Hamiltonian for the system is 
\begin{equation}
H_1 = \hbar \frac{K}{2} (\left| 1\right\rangle \left\langle 1\right| -\left|
0\right\rangle \left\langle 0\right| )+ \hbar \Omega \left(e^{-i\varphi }\sigma
^{+}+e^{i\varphi }\sigma ^{-}\right),
\label{eq1}
\end{equation}
where $\hbar $ is the reduced Planck constant, $\sigma ^{+}$ 
($\sigma ^{-}$) 
is the qubit raising (lowering) operator, 
$K$ is a function of time representing the qubit frequency
fluctuation due to the presence of noise, and $\Omega $ and $\varphi $ are the
Rabi frequency and phase of the driving field, respectively. The continuous drive creates
two dressed states, $\left| +\right\rangle =\frac 1{\sqrt{2}}(\left|
0\right\rangle +e^{i\varphi }\left| 1\right\rangle )\ $and $\left|
-\right\rangle =\frac 1{\sqrt{2}}(\left| 0\right\rangle -e^{i\varphi }\left|
1\right\rangle )$, that are separated by an energy gap $2\hbar\Omega $. In the basis $%
\left\{ \left| +\right\rangle ,\left| -\right\rangle \right\} $, we can
rewrite $H_1$ as 
\begin{equation}
H_1 = -\hbar \frac{K}{2} \left(\left| +\right\rangle \left\langle -\right| +\left|
-\right\rangle \left\langle +\right| \right)+ \hbar \Omega \left(\left| +\right\rangle
\left\langle +\right| -\left| -\right\rangle \left\langle -\right| \right).
\label{eq2}
\end{equation}
We assume that the noise correlation time is much longer than the dynamical
time scale $\Omega ^{-1}$, so that the fluctuation is adiabatic with respect to
the Rabi oscillation. With this assumption and under the condition $\left|
K\right| \ll \Omega $, the noise does not induce transitions between the two
dressed states; 
instead it leads to energy shifts of the dressed states. The resulting
effective Hamiltonian is 
\begin{equation}
H_{1,\textrm{eff}} = \hbar \left(\frac{K^2}{8\Omega }+\Omega \right) \left(\left| +\right\rangle
\left\langle +\right| -\left| -\right\rangle \left\langle -\right| \right).
\end{equation}
The effective qubit-environment
coupling is reduced by a factor $\frac K{4\Omega }$, which therefore produces a
significantly reduced phase difference (phase error), \mbox{$\epsilon =-\int_{t_1}^{t_2}$ $\frac{K^2}{4\Omega}dt$}, 
between 
$\left| +\right\rangle $ and 
$\left| -\right\rangle $ during the time interval [$t_1$, $t_2$]. 
In addition, the part of $\epsilon$ originating from the slowly-varying component of $K$ 
can be further reduced by employing a spin-echo like technique with a $\pi $-phase
shift of the driving field in the middle of [$t_1$, $t_2$]:
The integrand in $\epsilon$ reverses sign due to the $\pi $-phase
shift, and as a consequence, for a slow varying $K$, the phase error
accumulated during the first half of [$t_1$, $t_2$] exactly cancels out that of the second half. 
This single-qubit dynamical decoupling procedure, which is
analogous to the rotary echo technique originally demonstrated with nuclear
spins~\cite{Solomon1959PRL}, is named 1Q-DD here and below in contrast to the two-qubit case.

To incorporate the DD procedure into the implementation of
a two-qubit controlled phase gate, we consider the system consisting of two
on-resonantly coupled qubits, Q$_1$ and Q$_2$, each resonantly driven by a classical field
for the gate operation as well as for DD protection, named 2Q-DD here and below.
In the interaction picture (qubit reference frame), the Hamiltonian is 
\begin{equation}
H_2 = \hbar \left( \lambda \sigma _1^{+}\sigma _2^{-}+\sum_{j=1,2}\Omega
_je^{-i\varphi _j}\sigma _j^{+}\right) + \textrm{h.c.},
\label{eq4}
\end{equation}
where h.c. stands for Hermitian conjugate, $\lambda $ is the qubit-qubit
excitation swapping rate, and $\Omega _j$ and $\varphi _j$ are the Rabi
frequency and phase, respectively, of the drive applied to Q$_j$, which produces
two dressed states $\left| +_{\varphi _j,j}\right\rangle =\frac 1{\sqrt{2}}%
(\left| 0_j\right\rangle +e^{i\varphi _j}\left| 1_j\right\rangle )\ $and $%
\left| -_{\varphi _j,j}\right\rangle =\frac 1{\sqrt{2}}(\left|
0_j\right\rangle -e^{i\varphi _j}\left| 1_j\right\rangle )$. For simplicity,
here the qubit frequency fluctuations, described by $K$ in Eq.~(\ref{eq1}), are not included,
whose effects are suppressed by application of the continuous drives as discussed above. Under
the conditions $\varphi_1=\varphi_2=\varphi$ and $\left| \Omega _1-\Omega _2\right| \gg \left|
\lambda \right| $, the qubits cannot undergo transitions between different
dressed states due to the energy gaps produced by the drives. Then the system dynamics can be approximately described by the effective Hamiltonian (see Supplemental Material)
\begin{equation}
H_{2,\textrm{eff}} = \frac{1}{2} \hbar \lambda  S_{z,\varphi,1}S_{z,\varphi,2}+ \hbar \sum_{j=1,2}\Omega _jS_{z,\varphi,j},
\label{eq5}
\end{equation}
where $S_{z,\varphi,j}=\left|
+_{\varphi,j}\right\rangle \left\langle +_{\varphi,j}\right| -\left|
-_{\varphi,j}\right\rangle \left\langle -_{\varphi,j}\right|$. The first term results in conditional phase shift in the dressed state basis, which is
responsible for the two-qubit entangling gate. 
The terms in summation are for single-qubit rotations, 
which produce null effect if $\varphi _1$ and $\varphi _2$
of both driving fields are reversed by $\pi$ right in the middle of the two-qubit interaction time $\tau$. 
Then the two-qubit dressed states evolve as
\begin{eqnarray}
\left| +_{\varphi,1}\right\rangle \left| +_{\varphi ,2}\right\rangle
&\longrightarrow &e^{i\theta }\left| +_{\varphi ,1}\right\rangle \left|
+_{\varphi,2}\right\rangle ,  \nonumber \\
\left| +_{\varphi,1}\right\rangle \left| -_{\varphi ,2}\right\rangle
&\longrightarrow &e^{-i\theta }\left| +_{\varphi ,1}\right\rangle \left|
-_{\varphi,2}\right\rangle ,  \nonumber \\
\left| -_{\varphi,1}\right\rangle \left| +_{\varphi ,2}\right\rangle
&\longrightarrow &e^{-i\theta }\left| -_{\varphi ,1}\right\rangle \left|
+_{\varphi,2}\right\rangle ,  \nonumber \\
\left| -_{\varphi,1}\right\rangle \left| -_{\varphi ,2}\right\rangle
&\longrightarrow &e^{i\theta }\left| -_{\varphi ,1}\right\rangle \left|
-_{\varphi,2}\right\rangle ,
\label{eq6}
\end{eqnarray}where $\theta =-\frac 12 \lambda \tau$. With $\theta=\pi/4$ the state evolution in Eq.~(\ref{eq6}) naturally yields a conditional phase
gate in the dressed state basis.
For simplicity, we take $\varphi=0$, and then the corresponding unitary matrix in the two-qubit computational basis \{$|00\rangle$, $|01\rangle$, $|10\rangle$, $|11\rangle$\} is given by
\begin{equation}
\label{Um}
U_\textrm{phase} = \frac{1}{\sqrt{2}} \left(\begin{matrix}
 1 & 0 & 0 & i \\
 0 & 1 & i & 0 \\
 0 & i & 1 & 0 \\
 i & 0 & 0 & 1
\end{matrix}\right).
\end{equation}
Together with single-qubit rotations $e^{i\theta S_{z,\varphi,j}}$
which incur extra phase factors $e^{i2\theta}$ between $|+_{\varphi,j}\rangle$ and $|-_{\varphi,j}\rangle$ for both qubits, 
this operation is equivalent to a controlled phase gate, 
where the phase of the two-qubit system is effectively shifted by 
$4\theta $ if and only if the system is in the state $\left| +_{\varphi
,1}\right\rangle \left| +_{\varphi,2}\right\rangle $. 
We note that the dressed state phase gate cannot be produced without the continuous drives.

\begin{figure}[t]
  \centering
  \includegraphics[width=3.1in,clip=True]{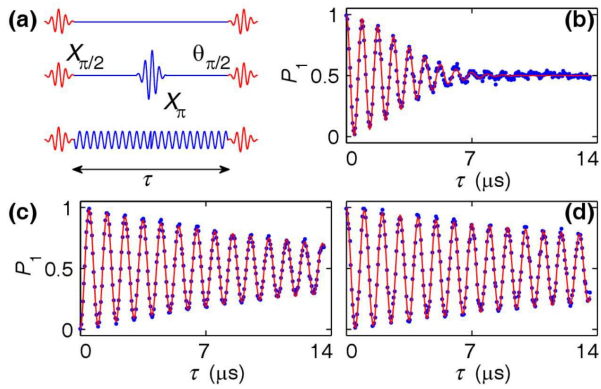}\caption{\label{fig1}\footnotesize{(a) Pulse sequences for Ramsey interference under free decay (top), spin-echo (middle), and
	1Q-DD (bottom).
	The $X_{\pi /2}$ ($X_\pi$) pulse rotates the qubit by $\pi/2$ ($\pi$) around x axis;
	The $\theta_{\pi /2}$ pulse rotates the qubit by $\pi/2$ around
 	the axis with an angle $\theta =\omega _R\tau $ to x axis on xy plane, 
	where $\tau $ is the interval between the two $\pi/2$ rotation pulses 
	and $\omega _R/2\pi$ is adjusted to $\approx 1.0$ MHz in this experiment.
	The center long wave in the bottom panel illustrates the continuous drive with 
	a Rabi frequency of $\Omega \approx 3.6$~MHz whose phase is inverted at $\tau /2$.
	Q$_1$'s $|1\rangle$-state probabilities after readout corrections (dots) as functions
	of $\tau$ are shown for free decay in (b), spin-echo in (c), and 1Q-DD in (d).
	Lines are fits according to the envelope decay of $P_1^\textrm{env} =
	0.5+0.5\exp \left[ -\tau /\left( 2T_1\right) -\left( \tau /T_2^{\ast}\right) ^2\right]$ for (b)
	and $0.5+0.5\exp\left(-\tau/T_\textrm{d}\right)$ for (c) and (d), 
	in combination with a fast oscillation term $\cos{\left( \omega _R \tau+\varphi_0 \right)}$.
	The pure dephasing time $T_\phi$ is estimated using $1/T_\textrm{d} = 1/(2T_1) + 1/T_\phi$. 
	See Supplemental Material for more data on qubit coherence.
	}}
\end{figure}

\begin{figure}[b]
  \centering
  \includegraphics[width=3.4in,clip=True]{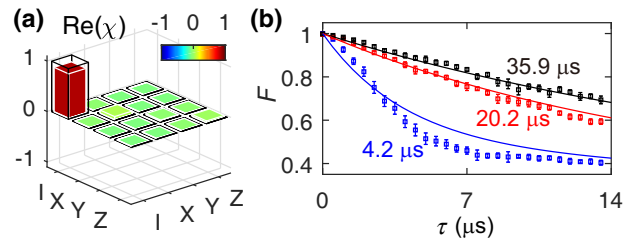}
  \caption{\label{fig2}\footnotesize{(a) Real components of the experimental QPT matrix $\chi_\textrm{exp}$ (solid bars) in comparison with the ideal 
	$\chi_\textrm{id}$ (identity matrix shown as black frames) for
	characterizing the the storage integrity in Q$_1$ for a time of 5~$\mu$s using the 1Q-DD procedure,
	where $I$ is the identity and \{$X$, $Y$, $Z$\} are the Pauli operators
  \{$\sigma_x$, $\sigma_y$, $\sigma_z$\} defined in the single-qubit computational basis \{$|0\rangle$, $|1\rangle$\}~\cite{Zheng2017}.
	All imaginary components (data not shown) are measured to be no higher than 0.0028.
    (b)	Fidelity $F = \textrm{tr}\left(\chi_\textrm{exp}\chi_\textrm{id}\right)$
	as a function of the storage time $\tau$ using the 1Q-DD procedure (black dots). Also shown are those with spin-echo (red dots)
	and free decay (blue dots). Lines are calculated QPT matrix fidelities using the Lindblad master equation for a free-decay process,
	with $T_1 = 31.6~\mu$s and the listed values for the pure dephasing time $T_\phi$.
	For the free-decay process, the experimental $F$ versus $\tau$ (blue dots) has  
	a Gaussian envelope due to the $1/f$ dephasing noise.}}
\end{figure}

\begin{figure}[t]
	\centering
	\includegraphics[width=3.4in,clip=True]{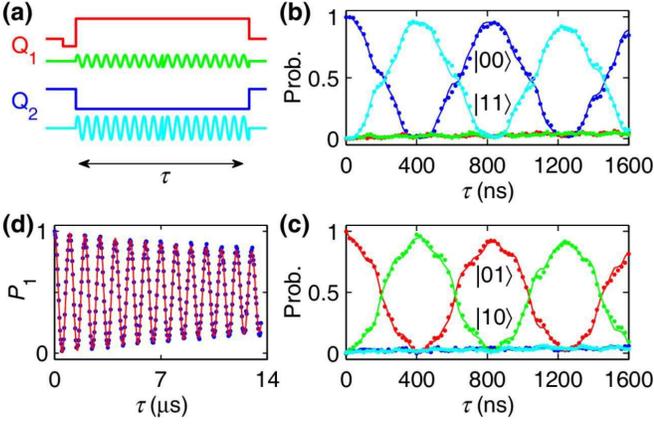}\caption{\label{fig3}\footnotesize{
	(a) Pulse sequences (square Z bias and sinusoid microwave for each qubit) for the two-qubit phase gate operation with 2Q-DD, where
	both qubits are dynamically tuned on resonance at the gate point for interaction with the big square pulses. 
	Two phase adjustments involving three parameters are critical to validate Eq.~(\ref{eq5}): 
    A small square pulse with an adjustable amplitude, which is 15 ns in width, is applied to
    Q$_1$ right before the square pulse to align its x axis of the Bloch sphere to that of Q$_2$'s in the rotating
		frame of the on-resonance frequency; 
    the initial phase $\varphi_j$ of each drive $\Omega_je^{-i\varphi_j}$, which is latter inverted at $\tau/2$, 
		is aligned to the x axis of each qubit, so that $\varphi_1=\varphi_2$ for the maximum coupling strength. 
	Here $\Omega_1/2\pi\approx3.6$~MHz and $\Omega_2/2\pi\approx6.9$~MHz.
	(b) Experimental data (dots) and numerical simulations (solid lines) for the evolution of populations of the four computational
	states, $\left| 00\right\rangle $ (blue),  $\left| 01\right\rangle $ (red), $\left| 10\right\rangle $ (green), 
	and $\left| 11\right\rangle $ (cyan), with the initial input state in $\left| 00\right\rangle $.
	For numerical simulations, we use the Lindblad master equation and include the microwave crosstalk effect~\cite{CSong2017a}, 
	where the pure dephasing times of the two qubits are set to the $T_\phi$ values obtained in the 2Q-DD procedures as
	exemplified in {\bf d}.
	(c) Similar data as in \textbf{b} with the initial input state in $\left| 01\right\rangle $.
	The nonsinusoidal effects in {\bf b} and {\bf c} are due to the experimental nonideality 
	with respect to the requirement $\left|\Omega_1 -\Omega_2\right| \gg \left|\lambda\right|$, 
	which leads to nonvanishing transition probabilities between different dressed states.
	(d) Ramsey interference data of Q$_1$, 
	while Q$_2$ is maintained in the dressed state $|+_{\varphi,2}\rangle$ throughout the 2Q-DD procedure.
	Q$_1$'s pure dephasing time during the 2Q-DD procedure is estimated 
	using the Ramsey decay envelope with the equations provided in the caption of Fig.~1.
	All probability data are corrected for readout errors~\cite{Zheng2017}.
	}}
\end{figure}

We benchmark the dephasing-insensitive scheme with a circuit quantum electrodynamics (QED)
architecture consisting of multiple frequency-tunable transmon qubits
connected by a fixed-frequency ($\omega _\textrm{r}/2\pi \approx 5.795$~GHz) resonator, which is
used to mediate the qubit-qubit interaction~\cite{CSong2017a} required for the two-qubit gate.
Here we take a representative qubit, Q$_1$, with an energy relaxation time $T_1\approx31.6$ $ \mu$s operating at 5.643~GHz,
the gate point that is 152 MHz below $\omega _\textrm{r}/2\pi$, as an example to illustrate the
suppression of dephasing using our 1Q-DD procedure. We first 
decouple this qubit from other qubits by tuning them far off-resonance,
and initialize it to $|0\rangle$ by idling for more than 200~$\mu$s. 
Figure 1 shows the qubit $|1\rangle$-state probabilities $P_1$ as functions of the
interval $\tau$ between two Ramsey pulses under free decay, spin-echo \cite{Hahn1950PR}, and 1Q-DD,
with the corresponding pulse sequences shown in Fig.~1(a). 
The Gaussian dephasing time $T_2^{*}$~\cite{Averin2016PRL} estimated from Fig.~1(b) is $\approx 4.2~\mu$s,
and dramatic improvements of phase coherence are observed in Figs.~1(b) and (c).
Since the interferometry data are limited to $\tau < 14$~$\mu$s due to our hardware constraints
and the fitted dephasing time in Figs.~1(c) and (d) could be much longer,
here and below we introduce a single exponential term with a time constant $T_\textrm{d}$ to
describe the combined energy decay and dephasing impact for more confidence of the fitted value.
$T_\textrm{d} \approx 15.3 \,\mu$s in Fig.~1(c) and $\approx 22.9\,\mu$s in Fig.~1(d),
which correspond to a spin-echo $T_\phi$ of 20.2~$\mu$s and a 1Q-DD $T_\phi$ of 35.9~$\mu$s, respectively.
We note that, in a recent experiment \cite{Gustavsson2012PRL}, the rotary echo
technique was used to mitigate the dephasing due to slow fluctuations in the
drive amplitude.

More convincingly, we can store quantum information in the qubit
and examine the storage integrity using the single-qubit quantum process tomography (QPT)
to witness the effectiveness of our dephasing-insensitive 1Q-DD scheme.
The QPT is achieved by preparing a set of 6 
input states via the single-qubit gates \{$I$, $\pm X_{\pi /2}$, $\pm Y_{\pi /2}$, $X_\pi$\}, 
and measuring the resulting density matrices and
those after the storage through quantum state tomography~\cite{Zheng2017}.
An example of the process matrix $\chi_\textrm{exp}$ after a storage of 5~$\mu$s is shown in Fig.~2(a),
and the $\chi_\textrm{exp}$ fidelity $F$ 
as a function of the storage time
$\tau$ is shown in Fig.~2(b). 
For comparison, we also measure the $\chi_\textrm{exp}$ matrices for the cases under free decay 
and with the spin-echo technique
during the storage, both yielding lower fidelities.
Numerical simulations (lines) using $T_1$ and the listed values for $T_\phi$
in Fig.~2(b) further confirm that the 1Q-DD scheme is efficient in protecting the qubit from dephasing. 

\begin{figure}[t]
	\centering
	\includegraphics[width=3.4in,clip=True]{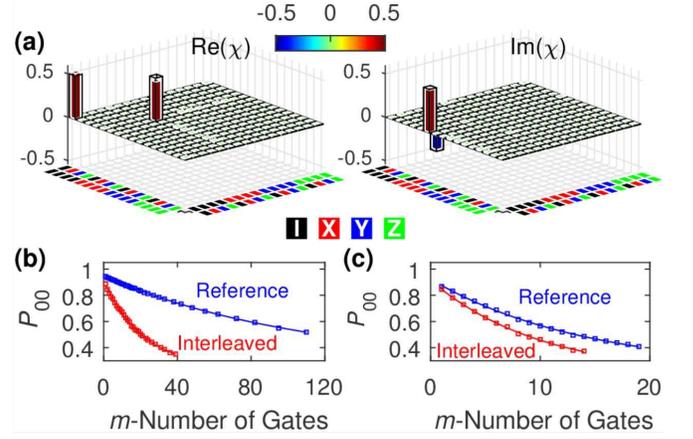}\label{fig4}
	\caption{\footnotesize{(a) Real and imaginary components of the experimental $\chi_{\rm{exp}}$ 
	(solid bars) and the ideal $\chi_{\rm{id}}$ (black frames)
	for the two-qubit $U_\textrm{phase}$ gate. 
	The color code for Pauli basis \{$I$, $X$, $Y$, $Z$\} is shown at the bottom. 
	$\chi_{\rm{exp}}$ has a fidelity of 0.9708 $\pm~0.0029$ and, for the 36 output states,
	the $|2\rangle$-state occupation probabilities of each qubit are no higher than 0.0046.
	(b) RB by inserting $U_\textrm{phase}$ between random single-qubit Pauli gates, yielding the gate fidelity of 0.9732~$\pm$ 0.0012.
	Plotted are the reference curve showing the uncorrected $|00\rangle$-state probability 
	after a series of random single-qubit Pauli gates (blue) 
	and the curve with $U_\textrm{phase}$ interleaved (red).
	(c) RB by inserting $U_\textrm{phase}$ between random Clifford gates, yielding the gate fidelity of $0.9781 \pm 0.0068$.
	The error bars, smaller than the size of the dots, are omitted to improve visual clarity.
	}}
\end{figure}

Now we turn to the 2Q-DD scheme for realizing the two-qubit entangling gate 
$U_\textrm{phase}$ by adding the second qubit, 
Q$_5$ in Ref.~\cite{CSong2017a} with $T_1 \approx 19.7$~$\mu$s, 
which is re-labeled as Q$_2$ here and below for the clarity of the presentation. 
Q$_1$ and Q$_2$ both physically connect to the central bus resonator 
with the coupling strengths of $g_1/2\pi \approx 14.2$~MHz and $g_2/2\pi \approx 15.2$~MHz, respectively, 
and there is no direct coupling between these two qubits when they are far detuned in frequency.
We can switch them on-resonance by tuning their frequencies to the same value which is detuned from the resonator frequency.
With this setting, these two qubits are directly 
coupled through virtual photon exchange mediated by the resonator (see Ref.~\cite{SBZheng2000PRL} and Supplemental Material). 
For any specific measurement with the pulse sequence shown in Fig. 3(a), 
we first initialize Q$_1$ and Q$_2$ at 5.613 and 5.673 GHz, respectively,
by creating any two-qubit product state $|\Psi_0\rangle$ using single-qubit rotations, while all other qubits are far detuned. 
We then apply square Z pulses to both qubits, tuning them on-resonance, so that these two qubits
are red-detuned from the resonator by the same amount of 152 MHz, the gate point. 
When the resonator is initially in the vacuum state, the
effective qubit-qubit coupling in Eq.~(\ref{eq4}) is measured to 
be $\lambda/2\pi \approx -1.2$ MHz with the two-qubit population swap process.
In Figs. 3(b) and (c), we present the measured (dots) and simulated (lines) 
two-qubit populations as functions of the gate duration, in the presence of the external driving fields, for the
initial states of $|\Psi_0\rangle = \left| 00\right\rangle$ and 
$\left| 01\right\rangle $, respectively. As expected, for the input state $\left| 00\right\rangle$, 
the states $\left| 00\right\rangle $ and $\left| 11\right\rangle $ periodically exchange
populations, while $\left| 01\right\rangle $ and $\left| 10\right\rangle $
are almost unpopulated throughout the gate duration; for the input state 
$\left| 01\right\rangle $, the observed anticorrelation between the
populations of $\left| 10\right\rangle $ and of $\left| 10\right\rangle $
are also in good agreement with theoretical predictions. 
We note that in the Lindblad master equation simulations~\cite{CSong2017a} in Figs.~3(b) and (c) 
the pure dephasing times $T_\phi$ are set to those obtained using the interferometry data exemplified in Fig.~3(d).
As shown in Fig.~3(d), with both qubits at the gate point and Q$_2$ initialized in the dressed state $|+_{\varphi, 2}\rangle$, 
the Ramsey interference pattern of Q$_1$ with 2Q-DD 
yields $T_\textrm{d}\approx 32.2~\mu$s and $T_\phi \approx 65.6~\mu$s.
$T_\phi$ of the 2Q-DD procedure is indeed longer than that of 1Q-DD, which has been verified previously~\cite{CSong2017a,Xu2018}.

At $\tau\approx 200$~ns, the Q$_1$-Q$_2$ interaction with 2Q-DD realizes $U_\textrm{phase}$. 
To characterize this gate, we perform the two-qubit QPT by preparing a full set of 36 distinct
input states through the two-qubit gates  \{$I$, $\pm X_{\pi /2}$, 
$\pm Y_{\pi /2}$, $X_\pi$\}$_1\otimes$\{$I$, $\pm X_{\pi /2}$, $\pm Y_{\pi /2}$, $X_\pi$\}$_2$. We present the
ideal and experimental process matrices $\chi _{\textrm{id}}$ and $%
\chi _\textrm{exp }$ in Fig.~4(a), finding a gate fidelity of $0.9708\pm0.0029$. 
Randomized benchmarkings (RBs) performed by inserting $U_\textrm{phase}$ between random Pauli- and 
Clifford-based gates are shown in Figs.~4(b) and (c), which yield consistent gate 
fidelity values of $0.9732 \pm 0.0012$ and $0.9781 \pm 0.0068$, respectively. 

Numerical simulations suggest that our measured $U_\textrm{phase}$ gate fidelity above 0.97
is consistent with the qubit $T_\phi$ values obtained in the 2Q-DD procedure,
where approximately 30\% of the total error are due to $T_1$, 
10\% due to $T_\phi$, and 60\% due to limited anharmonicites of the qubits.
As the dephasing is effectively suppressed by continuous drives (see Supplemental Material 
for more experimental data and discussions), 
the resulting error can be further reduced by using devices with
longer qubit $T_1$, stronger qubit-qubit coupling, 
and larger qubit anharmonicity \cite{Lv2012}. 
We perform a numerical simulation with the parameters
$\lambda/2\pi =  3$ MHz, 
$\Omega _1/2\pi = 12.4$ MHz, 
$\Omega _2/2\pi = 24.6$ MHz, 
$T_1=50$ $\mu $s, a pure dephasing time of $100~\mu$s, 
and an anharmonicity of 0.5 GHz \cite{You2007,Yan2016ncomms}, and find a gate
fidelity above 99.3\% that approaches the fault-tolerant threshold~\cite{Barends2014nature,Sheldon2016}.
Here $\Omega_1$ and $\Omega_2$ are reasonably low, which help to minimize the crosstalk issue~(see Supplemental Material)
so that the two-qubit gate with the dephasing-insensitive 2Q-DD scheme 
is applicable in the all-to-all connected circuit.

In conclusion, we have proposed and demonstrated a dephasing-insensitive
method for storing and processing quantum information in an all-to-all superconducting
circuit. The results show that the continuous, but weak, driving fields with phase reversals 
at the middle of the storage time can
dynamically decouple the qubits from dephasing noise.
We integrate this DD procedure with a
two-qubit quantum logic gate by applying weak continuous drives, which not only
enable the high-fidelity entangling gate for two coupled qubits
without invoking a non-computational state, but also protect the qubits from
dephasing noise during the gate operation.
\\

\noindent{\textbf{Acknowledgments}}.
This work was supported by the National Basic Research Program of China
under Grants No. 2014CB921201 and No. 2014CB921401, the National Natural Science Foundations
of China under Grants No. 11674060, No. 11434008, No. 11574380, and No. 11374344,
and the Fundamental Research Funds for the Central Universities
of China (Grant No. 2016XZZX002-01).
Devices were made at the Nanofabrication
Facilities at Institute of Physics in Beijing, University
of Science and Technology of China in Hefei, and National
Center for Nanoscience and Technology in Beijing.

\renewcommand\thefigure{S\arabic{figure}}
\renewcommand\theequation{S\arabic{equation}}
\renewcommand\thetable{S\arabic{table}}

\setcounter{figure}{0}
\setcounter{equation}{0}
\setcounter{table}{0}

\renewcommand\thefigure{S\arabic{figure}}
\renewcommand\theequation{S\arabic{equation}}
\renewcommand\thetable{S\arabic{table}}
\renewcommand\thesection{\arabic{section}}
\renewcommand\thesubsection{\thesection.\arabic{subsection}}

\begin{center}
{\noindent {\bf Supplementary Material for ``Dephasing-insensitive quantum information storage and processing with superconducting qubits''}}
\end{center}

\section{Effective qubit-qubit interaction under a strong drive}
In the basis formed by the dressed states $\left| +_{\varphi
_j,j}\right\rangle =\frac 1{\sqrt{2}}(\left| 0_j\right\rangle +e^{i\varphi
_j}\left| 1_j\right\rangle )\ $and $\left| -_{\varphi _j,j}\right\rangle =%
\frac 1{\sqrt{2}}(\left| 0_j\right\rangle -e^{i\varphi _j}\left|
1_j\right\rangle )$, the qubit flip operators $\sigma _j^{+}=\left|
0_j\right\rangle \left\langle 1_j\right| $ and $\sigma _j^{-}=\left|
1_j\right\rangle \left\langle 0_j\right| $ can be expressed, respectively, as 
\begin{eqnarray*}
\sigma _j^{+} &=&\frac 12e^{i\varphi _j}\left( S_{z,\varphi _j,j}-S_{\varphi
_j,j}^{+}+S_{\varphi _j,j}^{-}\right) , \\
\sigma _j^{-} &=&\frac 12e^{-i\varphi _j}\left( S_{z,\varphi
_j,j}+S_{\varphi _j,j}^{+}-S_{\varphi _j,j}^{-}\right) ,
\end{eqnarray*}
where $S_{z,\varphi _j,j}=\left| +_{\varphi ,j}\right\rangle \left\langle
+_{\varphi ,j}\right| -\left| -_{\varphi ,j}\right\rangle \left\langle
-_{\varphi ,j}\right| $, $S_{\varphi _j,j}^{+}=\left| +_{\varphi
_j,j}\right\rangle \left\langle -_{\varphi _j,j}\right| $ and $S_{\varphi
_j,j}^{-}=\left| -_{\varphi _j,j}\right\rangle \left\langle +_{\varphi
_j,j}\right| $. With this, we can rewrite the Hamiltonian of Eq. (4) of the
main text as 
\begin{multline}
H_2 =\frac 14\hbar \lambda \left[ e^{i\alpha }\left( S_{z,\varphi
_1,1}+S_{\varphi _1,1}^{+}-S_{\varphi _1,1}^{-}\right) 
 \left( S_{z,\varphi_2,2}+\right. \right. \\
\left. \left. S_{\varphi _2,2}^{-}-S_{\varphi _2,2}^{+}\right) + \textrm{h.c.} \right] +\hbar \sum_{j=1,2}\Omega _jS_{z,\varphi _j,j},
\end{multline}
where $\alpha =\varphi _1-\varphi _2$. The energy gap between the dressed
states $\left| +_{\varphi _j,j}\right\rangle $ and $\left| -_{\varphi
_j,j}\right\rangle $ produced by the driving term $\hbar \Omega
_jS_{z,\varphi _j,j}$ is $2\hbar \Omega _j$. Under the condition $\Omega
_j,\left| \Omega _1-\Omega _2\right| \gg \left| \lambda \right| $, the
qubits cannot undergo transitions between different dressed states due to
the large detunings, so that the terms containing $S_{\varphi _j,j}^{\pm }$
can be neglected. We note that, when $\alpha =0$ or $\pi $, the strong
driving condition can be somewhat loosened. We here take $\varphi _1=\varphi
_2=\varphi $ as an example. With this setting, $H_2$ reduces to 
\begin{multline}
H_2 = \frac 12\hbar \lambda \left( S_{z,\varphi ,1}S_{z,\varphi
,2}+S_{\varphi ,1}^{+}S_{\varphi ,2}^{-}+S_{\varphi ,1}^{-}S_{\varphi
,2}^{+}- \right. \\
\left. S_{\varphi ,1}^{+}S_{\varphi ,2}^{+}-S_{\varphi ,1}^{-}S_{\varphi
,2}^{-}\right)   
+\hbar \sum_{j=1,2}\Omega _jS_{z,\varphi ,j}.
\end{multline}
In this case, the second and third terms in the parentheses correspond to
the coupling between the two dressed states $\left| +_{\varphi
,1}\right\rangle \left| -_{\varphi ,2}\right\rangle $ and $\left| -_{\varphi
,1}\right\rangle \left| +_{\varphi ,2}\right\rangle $ with an energy gap $%
2\hbar \left| \Omega _1-\Omega _2\right| $, while the last two terms in the
parentheses describe the coupling between $\left| +_{\varphi
,1}\right\rangle \left| +_{\varphi ,2}\right\rangle $ and $\left| -_{\varphi
,1}\right\rangle \left| -_{\varphi ,2}\right\rangle $ with an energy gap $%
2\hbar \left( \Omega _1+\Omega _2\right) $. When $\left| \Omega _1-\Omega
_2\right| \gg \left| \lambda \right| $ , these couplings do not induce
transitions between different dressed states due to large detunings and thus
can be discarded. Then the system Hamiltonian reduces to Eq. (5) of the main
text.

\section{Selective resonator-induced qubit-qubit interaction}

When Q$_1$ and Q$_2$ are red-detuned from the resonator by the same amount 
$\Delta =\omega _r-\omega _q$ that is much larger than the qubit-resonator
couplings $g_1$ and $g_2$, they are coupled through virtual photon exchange
mediated by the resonator~\cite{SBZheng2000PRL}. If each of the other qubits is far
off-resonance with the resonator, Q$_1$, and Q$_2$, the Q$_1$-Q$_2$ dynamics
is not affected by the other qubits and the photon number of the
resonator remains unchanged during the interaction. When the resonator is
initially in the vacuum state, the effective Hamiltonian for Q$_1$ and Q$_2$
is given by~\cite{SBZheng2000PRL}
\begin{equation}
H_{\textrm{eff}}=H_{\textrm{eff}}^{j,k}+H_{\textrm{eff}}^0, 
\end{equation}
where 
\[
H_{\textrm{eff}}^{j,k}=\hbar \lambda \left( \sigma _1^{+}\sigma _2^{-}+\sigma
_1^{-}\sigma _2^{+}\right) , 
\]
\[
H_{\textrm{eff}}^0=-\hbar \left( \frac{g_1^2}\Delta \left| 1_1\right\rangle
\left\langle 1_1\right| +\frac{g_2^2}\Delta \left| 1_2\right\rangle
\left\langle 1_2\right| \right) , 
\]
and $\lambda =-g_1g_2/\Delta $. The vacuum-induced Stark shifts described by 
$H_{\textrm{eff}}^0$ can be compensated for by suitably adjusting the frequencies of Q$_1$ and Q$_2$, 
so that the effective Hamiltonian reduces to $H_{\textrm{eff}}^{j,k}$.  
The magnitude of the experimentally measured qubit-qubit coupling is
slightly smaller than $g_1g_2/\Delta $ for the existence of the direct
coupling~\cite{CSong2017a} between Q$_1$ and Q$_2$, which partly cancels out the
resonator-induced coupling whose sign is opposite to that of the direct
coupling.

\section{Device information}
The physical device used in our experiment is the same as that reported in Ref.~\cite{CSong2017a}, 
which consists of 10 frequency-tunable transmon qubits interconnected by a fixed-frequency central bus resonator $R$. 
The bus resonator $R$ is a superconducting half-wavelength coplanar waveguide resonator with 10 
side arms, and each arm capacitively couples to one qubit. 
Figure S1 shows a sketch of the device. More details on the device and 
the wiring configuration can be found in Supplemental Material of Ref.~\cite{CSong2017a}.
 
\begin{figure*}[h]
\includegraphics[width=11cm,clip=True]{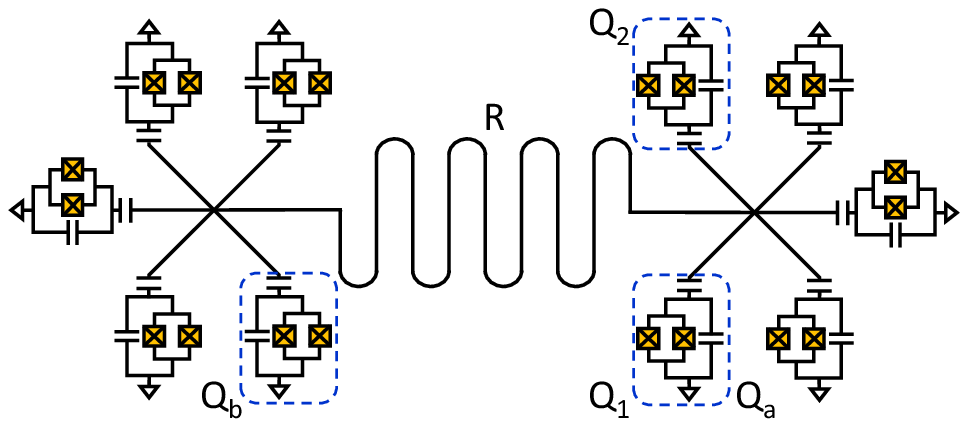}
\caption{Sketch of the device. Ten qubits capacitively couple to a central bus resonator $R$. 
Each qubit is composed of a SQUID loop with two Josephson junctions in parallel (orange squares with black crosses) 
and a shunted capacitance. The dashed-line boxes identify the qubits used in the experiment.
Q$_1$ and Q$_\textrm{a}$ refer to the same qubit but operated in different cooldowns.}
\end{figure*}

Owing to the all-to-all connectivity, arbitrarily selected two qubits can be coupled by detuning them 
from the bus resonator by the same amount in frequency. In the main text, we have described the experiment using Q$_1$ and Q$_2$ to implement the $U_{\rm{phase}}$ gate.
In this Supplemental Material, we provide additional data obtained in another experiment which was performed in a separate cooldown,
where qubit conditions changed so that we have to replace Q$_2$ by another qubit, Q$_{\rm{b}}$, and rename Q$_1$ as Q$_{\rm{a}}$
for the clarity of the presentation. Relevant qubit parameters at the interaction frequency for the $U_{\rm{phase}}$ gate are summarized in Table S1, with the experimental data shown in Fig.~S2.

\begin{table*}[htbp]
	 \setlength{\columnsep}{ 8pt} 
	\centering
	{\tabcolsep0.18in  \begin{tabular}{cccccc}
		\hline
		\hline
		     &$\omega_{I}/2\pi$ (GHz) &$T_1$ ($\mu$s) &$T_2^*$ ($\mu$s) &$g/2\pi$ (MHz) & $U_{\rm{phase}}$ gate length (ns) \\
		\hline     
		Q$_1$&\multirow{ 2}{*}{5.643}             &31.6           &4.2              &14.2 &\multirow{ 2}{*}{215.0}   \\
		Q$_2$&             &19.7           &3.4              &15.2 &     \\
		\hline
		Q$_{\rm{a}}$&\multirow{ 2}{*}{5.610}      &32.5           &$\approx$ 3.2       &14.2 &\multirow{ 2}{*}{210.7}   \\
		Q$_{\rm{b}}$&      &21.5           &3.5              &16.3 &     \\
		\hline
		\hline
	\end{tabular}}
	\caption{\label{Table 1} \footnotesize{Qubit and $U_{\rm{phase}}$ characteristics. $\omega_{I}$ is the interaction frequency for the $U_{\rm{phase}}$ gate.
	$T_1$ and $T_2^{*}$ are the single-qubit energy decay time and Gaussian dephasing time, respectively, for each qubit measured at $\omega_{I}$. 
	$g$ is the qubit-bus resonator coupling strength.}}	
\end{table*}

\section{Qubit pure dephasing times obtained in different procedures}

\begin{figure*}[h]
\includegraphics[width=16cm]{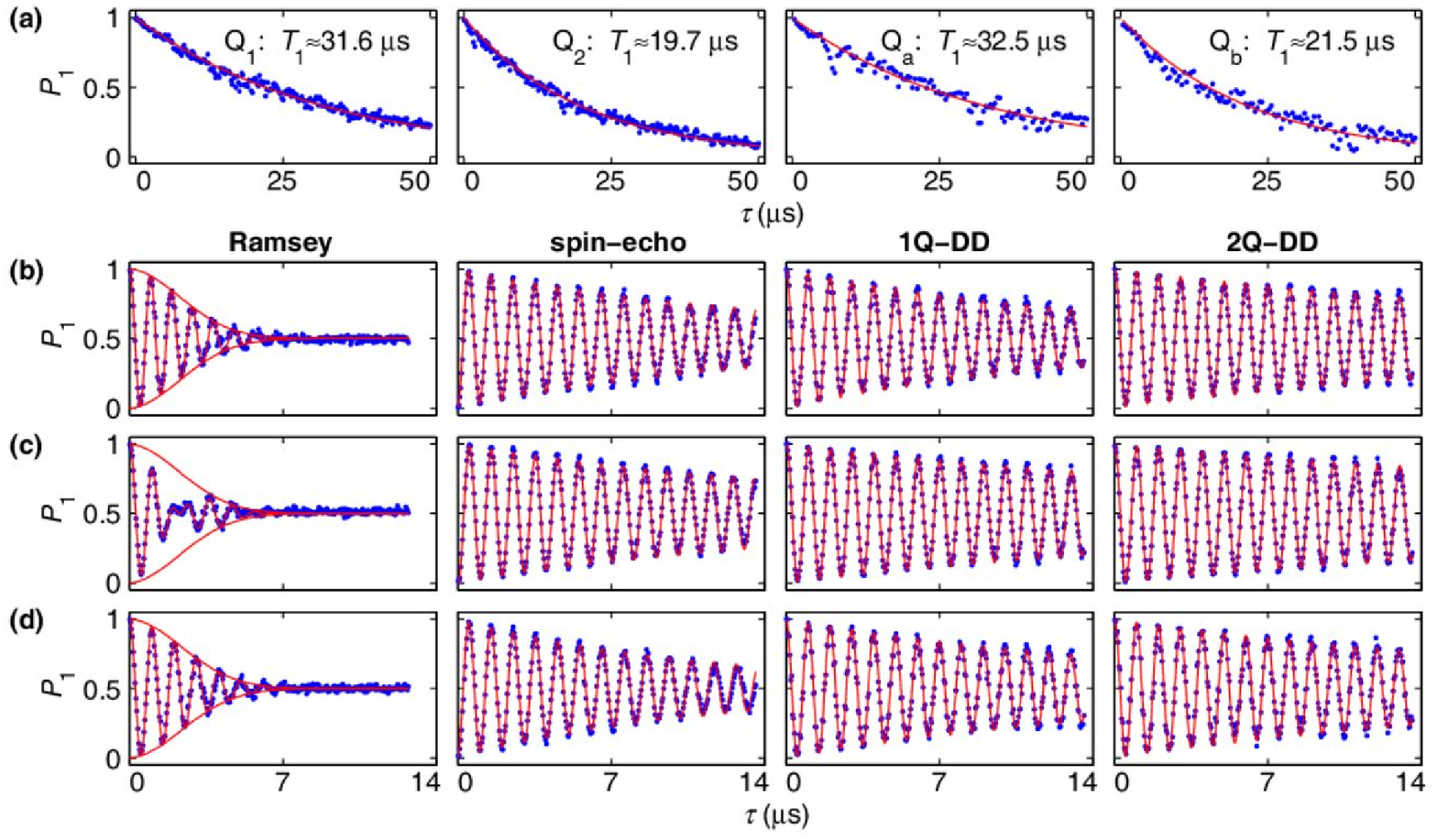}
\caption{\footnotesize{(a) Measurements of the qubits' energy decay times.  
The corrected $|1\rangle$-state probability of each qubit, $P_1$, is measured 
as a function of the decay time. Lines are fits according to $P_1=\exp(-\tau/T_1)$. 
Qubit interference patterns for Q$_{\rm{2}}$ (b), Q$_{\rm{a}}$ (c), and Q$_{\rm{b}}$ (d) obtained in different procedures as indicated (see the main text for the definitions of these procedures). 
All $P_1$ data are after readout corrections and lines with sinusoidal oscillations are fits according
the equations provided in the caption of Fig.~1 in the main text.
The Ramsey interference pattern of Q$_{\rm{a}}$ has a node at around 2~$\mu$s due to the perturbation of unknown type of noise, which
can be fixed by the noise-suppression schemes including spin-echo, 1Q-DD, and 2Q-DD.}}
\end{figure*}

The randomized benchmarking (RB) sequences can estimate higher gate fidelities 
due to their resemblance to DD sequences when low-frequency noise is present in the system \cite{Harrison2016PRA}.
To quantify such an effect, we have performed an additional experiment with Q$_{\rm{a}}$ and Q$_{\rm{b}}$,
where we bias both qubits to 185 MHz below the bus resonator for a $U_{\rm {phase}}$ gate length of 210.7 ns,  
and obtain the gate fidelity values of 0.9720 $\pm$ 0.0011 (QPT), 0.9784 $\pm$ 0.0012 (Pauli-based RB), 
and 0.9722 $\pm$ 0.0062 (Clifford-based RB). 

We define the idler gate where the target qubit is idled at $\omega_I$ for the time of the $U_{\rm {phase}}$ gate length 
while the other qubit is detuned. The idler gate has been characterized 
using the interleaved RB sequences as illustrated in Fig.~S3(a), 
whose fidelity is determined by the qubit $T_1$ and $T_{\phi}$ at $\omega_I$. 
The idler gate fidelity is 0.9843 for Q$_{\rm{a}}$ and 0.9884 for Q$_{\rm{b}}$ as shown in Fig.~S3(b) and (c).
We note that the idler-RB $T_{\phi}$, inferred from the idler gate fidelity, should indicate the enhancement of 
RB sequences due to their resemblance to DD sequences.  

It is seen from Table~S2 that idler-RB $T_{\phi}$ is significantly shorter 
than 1Q-DD $T_{\phi}$ and 2Q-DD $T_{\phi}$. In particular, the enhancement of $T_{\phi}$ due to RB sequences for Q$_{\rm{a}}$ 
is less obvious, indicating that RB sequences are less effective in suppressing the unknown type of noise.
Our numerical simulations show that 
the 2Q-DD $T_{\phi}$ values are consistent with the measured $U_{\rm {phase}}$ gate fidelity values on average, 
which cannot be obtained with the idler-RB $T_{\phi}$ values. With the idler-RB $T_{\phi}$ values listed in Table~S2,
our numerical simulations suggest that the $U_{\rm {phase}}$ gate fidelity maximizes at about 0.96.

\begin{table*}[htbp]
	 \setlength{\columnsep}{ 8pt} 
	\centering
	{\tabcolsep0.09 in  \begin{tabular}{ccccccc}
		\hline
		\hline
		     & spin-echo $T_{\phi}$ ($\mu$s)& 1Q-DD $T_{\phi}$ ($\mu$s)& 2Q-DD $T_{\phi}$ ($\mu$s) & idler-RB  $T_{\phi}$\\
		\hline     
		Q$_1$&   20.2& 35.9& 65.6& -   \\
		Q$_2$&   26.4& 24.7& 73.9& -   \\
		Q$_{\rm{a}}$& 28.7& 47.4& 78.7& 8.3 \\
		Q$_{\rm{b}}$& 16.6& 33.1& 45.0& 15.5\\
		\hline
		\hline
	\end{tabular}}
	\caption{\label{Table 1} \footnotesize{ 
	Various pure dephasing times $T_{\phi}$ estimated using  $1/T_{\rm d} = 1/(2 T_1) + 1/T_{\rm{\phi}}$, 
	where $T_d$ describes the decay rate of the interference envelope which is obtained using 
	the corresponding sequence as labeled: spin-echo and 1Q-DD correspond to the sequences 
	used in Figs. 1(c) and (d) of the main text, respectively; 2Q-DD corresponds the sequence 
	used in Fig.~3(d) of the main text. 
	To estimate the idler-RB $T_{\phi}$ we perform numerical simulations using the idler gate fidelity obtained by the interleaved RB method (see Fig.~S3).}}	
\end{table*}

\begin{figure*}[h]
\includegraphics[width=16cm]{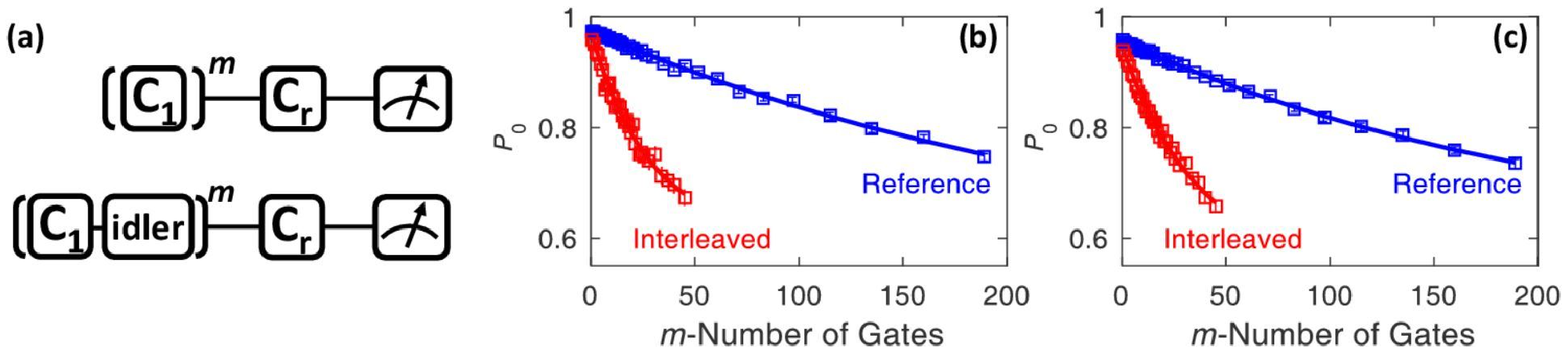}
\caption{\footnotesize{Idler gate RBs for Q$_{\rm{a}}$ and Q$_{\rm{b}}$. 
(a) Sequence diagram of the reference (top) and that interleaved with the idler gate (bottom), 
where C$_1$ is randomly chosen from the single-qubit Cliffords and C$_{\rm{r}}$ is the recovery gate. 
The idler gate length is 210.7 ns, the same as that of $U_{\rm {phase}}$.
Experimental RB results of the idler gate are shown in (b) for Q$_{\rm{a}}$ and (c) for Q$_{\rm{b}}$,
with gate fidelity values of 0.9843 and 0.9884, respectively. 
}}
\end{figure*}


\section{The crosstalk effect}
Crosstalk is an inevitable topic if the circuit continues to scale up. However, 
we expect that the hardware issue resulting in crosstalk can be significantly improved 
with better circuit designs and fabrication technologies. On the other hand, 
the performance of our dephasing-insensitive scheme is less sensitive to 
the crosstalk effect due to the following reasons.

For the two on-resonant qubits, Q$_1$ and Q$_2$, completing the $U_{\rm{phase}}$ gate, 
the mutual crosstalk only changes the effective amplitude and phase of the microwave 
applied to each qubit, where the overall gate dynamics remain the same and an optimal 
$U_{\rm{phase}}$ gate is achievable by adjusting the control parameters. 
With an additional qubit, e.g., Q$_{j}$, simultaneously driven but detuned by 
$\Delta_{j,\,1}$ from the on-resonance frequency of Q$_1$ and Q$_2$, the crosstalk effect 
results in weak and dispersive microwave drives on Q$_1$ and Q$_2$, which produce 
a Stark shift ($|\Omega^{\rm c}_{j}|^2/\Delta _{j,\,1})(|1\rangle\langle 1|-|0\rangle\langle 0|$) 
on either qubit (here $\Omega^{\rm c}_{j}$ is the dispersive drive strength through 
Q$_{j}$ on Q$_1$ or Q$_2$, and $\Omega^{\rm c}_{j} \ll \Delta_{j,\,1}$). 
We note that this energy-level shift only modifies the qubit frequency fluctuation referring to Eq. (1) of the main text, 
whose effect is suppressed by the corresponding dynamical decoupling procedure. 
Therefore the microwave crosstalk effect is negligible if qubits are properly separated 
in frequency. If more qubits are added into the system and the whole frequency band 
becomes crowded, we can choose to sequentially implement $U_{\rm{phase}}$ gates for some qubit pairs to avoid such crowding in frequency.

\end{document}